\documentclass{article}

\usepackage{arxiv}

\usepackage[utf8]{inputenc} 
\usepackage[T1]{fontenc}    
\usepackage{hyperref}       
\usepackage{url}            
\usepackage{booktabs}       
\usepackage{amsfonts}       
\usepackage{nicefrac}       
\usepackage{microtype}      
\usepackage{lipsum}
\usepackage{graphicx}
\usepackage{cite}
\usepackage{amsmath,amssymb,amsfonts}
\usepackage{algorithmic}
\usepackage{graphicx}
\usepackage{textcomp}
\usepackage{multirow}
\usepackage{amsmath}
\usepackage{amsfonts}
\usepackage{enumitem}
\usepackage{balance}
\usepackage{csquotes}
\usepackage{subcaption}
\usepackage{cleveref}
\usepackage{hyperref}
\hypersetup{breaklinks=true}
\usepackage{float}
\graphicspath{ {./images/} }

\begin{document}

\thispagestyle{empty}
{\Huge \textbf{IEEE Copyright Notice}}\\[1em]
{\large
\noindent
\textcopyright~2024 IEEE. Personal use of this material is permitted. Permission from IEEE must be obtained for all other uses, in any current or future media, including reprinting/republishing this material for advertising or promotional purposes, creating new collective works, for resale or redistribution to servers or lists, or reuse of any copyrighted component of this work in other works.

\vspace{1em}
\noindent
DOI: \href{https://doi.org/10.1109/ICISET62123.2024.10941254}{10.1109/ICISET62123.2024.10941254}
}
\newpage

\fancypagestyle{titlepage}{
  \fancyhf{}
  \fancyhead[C]{\footnotesize This work has been accepted for publication in 2024 4th Int. Conf. on Innovations in Science, Engineering and Technology(ICISET) 26-27 October 2024, Chittagong, Bangladesh.\\
  The final published version is available via IEEE Xplore.\\ DOI: \href{https://doi.org/10.1109/ICISET62123.2024.10941254}{10.1109/ICISET62123.2024.10941254}}
  \renewcommand{\headrulewidth}{0pt}
}

\title{A Comprehensive Analysis of COVID-19 Detection
Using Bangladeshi Data and Explainable AI}

\author{
 Shuvashis Sarker \\
  Department of Computer Science and Engineering\\
  Ahsanullah University of Science and Technology\\
  Dhaka,Bangladesh\\
  \texttt{shuvashisofficial@gmail.com}
}

\maketitle
\thispagestyle{titlepage}

\begin{abstract}
COVID-19 is a rapidly spreading and highly infectious virus which has triggered a global pandemic, profoundly affecting millions across the world. The pandemic has introduced unprecedented challenges in public health, economic stability, and societal structures, necessitating the implementation of extensive and multifaceted health interventions globally. It had a tremendous impact on Bangladesh by April 2024, with around 29,495 fatalities and more than 2 million confirmed cases. This study focuses on improving COVID-19 detection in CXR images by utilizing a dataset of 4,350 images from Bangladesh categorized into four classes: Normal, Lung-Opacity, COVID-19 and Viral-Pneumonia. ML, DL and TL models are employed with the VGG19 model achieving an impressive 98\% accuracy. LIME is used to explain model predictions, highlighting the regions and features influencing classification decisions. SMOTE is applied to address class imbalances. By providing insight into both correct and incorrect classifications, the study emphasizes the importance of XAI in enhancing the transparency and reliability of models, ultimately improving the effectiveness of detection from CXR images.
\end{abstract}


\section{Introduction}
COVID-19 resulting from the SARS-CoV-2 virus, was first identified in Wuhan, China, in early 2020 and officially designated as a pandemic by the WHO in January 2020. It primarily affects the respiratory system leading to conditions like viral-pneumonia and in severe cases requiring mechanical ventilation \cite{ibrahim2021pneumonia}. The pandemic resulted in an estimated 14.9 million deaths worldwide between January 2020 and December 2021\footnote{\url{https://www.who.int/news/item/05-05-2022-14.9-million-excess-deaths-were-\\associated-with-the-covid-19-pandemic-in-2020-and-2021}}. As of April 13, 2024, Bangladesh reported over 2,050,193 cases and 29,495 deaths \footnote{\url{https://dashboard.dghs.gov.bd/pages/covid19.php}}. While most cases were mild or moderate, older adults and those with preexisting conditions like heart disease or diabetes face a higher risk of severe illness.

Chest X-rays (CXR) are an easy and not too expensive screening method that can be used in many healthcare settings with limited resources\cite{vj2021deep}. ML techniques are able to evaluate X-rays or CT scans of the lungs to identify individuals with COVID-19 pneumonia. This method can be especially helpful in developing countries that don't have access to COVID-19 testing kits because it gives doctors another way to prove COVID-19 cases in big groups of people\cite{noak2022analysis,rasheed2021machine}.

Deep learning(DL) has gained significant interest because to its superior performance compared to standard models and its ability to extract features from CXR images with fewer training cycles. LSTM networks excel at rapidly detecting COVID-19 by  analyzing patterns in the training data. CNNs play a crucial role in extracting pertinent and noise-free information from pictures. The accuracy of this process improves as more layers are added to the network \cite{noak2022analysis, bhadra2020covid}. TL algorithms excel at detecting COVID-19 specially when applied to CXR images utilizing artificial neural networks(ANN) that have been trained on extensive datasets. These models have the ability to identify significant characteristics from various collections of images and demonstrate high performance even when there is a limited amount of COVID-19 specific data. This allows for quicker and more precise diagnosis. When adjusted and trained again, TL models become increasingly skilled at recognizing the distinctive patterns of COVID-19, greatly assisting in the response to the epidemic. These models enhance the speed and accuracy of COVID-19 detection, particularly in environments with limited resources \cite{talukder2024empowering, lakshmi2022pre}.
While existing solutions reliably detect COVID-19, their classification rates remain below the desired level and studies on Bangladeshi COVID-19 cases are limited. This study aims to address these challenges by:
\begin{enumerate}[label=\roman*]
    \item Oversampling a Bangladeshi COVID-19 dataset using SMOTE improves classification and recognition.
    \item Comprehensively studying advanced DL, TL and conventional ML approaches to find the best COVID-19 CXR model.
    \item XAI approaches are applied to enhance model transparency by revealing critical features and patterns in COVID-19 detection.
\end{enumerate}
This work aims to enhance the exactness and reliability of COVID-19 identification in CXR images, specifically utilizing data gathered from Bangladesh.

\section{Related Works}
\subsection{ML model developed to determine COVID-19}
ML has significantly enhanced COVID-19 detection by evaluating medical imaging and clinical data, improving both speed and accuracy in diagnosis. Integration with healthcare has provided effective ways to differentiate COVID-19 from other respiratory disorders. Elias P. Medeiros et al. \cite{medeiros2024applications} conducted a study using the COVID-19 Radiography Database, utilizing various texture descriptor extractors and supervised learning algorithms. They found that combining multiple extractors improved predictive performance, achieving an accuracy of 95.47\%. Similarly, Zaid Albataineh et al.\cite{albataineh2024covid} developed a segmentation method that effectively evaluates the impact of COVID-19 from CT scans by extracting texture features. The accuracy achieved for different stages is remarkable: 99.12\% for normal, 98.24\% for mild, 98.73\% for moderate and an impressive 99.9\% for severe cases. Tawsifur Rahman et al.\cite{rahman2023bio} created a multimodal system that can accurately predict the risk of mortality from COVID-19. This system utilized a stacking ensemble of machine learning classifiers and was trained on a dataset obtained from Italian hospitals. Their system, combining chest X-ray features and clinical biomarkers, achieved an accuracy of 89.03\%, surpassing models relying solely on X-ray images or clinical data by 6\%. Additionally, they devised a nomogram scoring technique for predicting mortality. Lastly, Jawad Rasheed et al.\cite{rasheed2021machine} presented a logistic regression classifier model to enhance learning and classification accuracy on a public CXR dataset. They used generative adversarial networks to extend the dataset to 500 images and employed PCA to reduce dimensionality. As a result, CNN and LR models achieved impressive accuracy rates of 95.2\% to 97.6\% without PCA and 97.6\% to 100\% with PCA. These studies collectively highlighted the potential of ML to revolutionize healthcare by offering rapid and reliable diagnostic tools.

\subsection{DL model developed to determine COVID-19}
The contributions of Deep Neural Networks (DNNs) in the field of medical imaging, particularly in the detection of COVID-19 from CXR have been significant. From several studies, we can see diverse approaches leveraging DNNs for enhanced accuracy and robustness in diagnosis. Saleh Albahli et al.\cite{albahli2020efficient} presented a novel DCNN model incorporating data augmentation to achieve a remarkable 98.08\% accuracy on a dataset of 7,346 images. Similarly, Enzo Tartaglione et al\cite{tartaglione2020unveiling} used a DCNN to overcome small data challenges, attaining the same high accuracy through six diverse datasets. Md. Rezaul Karim's\cite{9313304} Deep COVID Explainer with a dataset of 15,959 images employed DCNN and data augmentation to deliver impressive predictive values of 91.6\%, 92.45\%, and 96.12\% for normal, pneumonia, and COVID-19 cases respectively. Maya Pavlova\cite{pavlova2022covid} implemented a cutting-edge ensemble DCNN approach, leveraging a diverse multinational dataset to achieve impressive result values of 95.5\% and 97.0\% respectively. Sivaramakrishnan Rajaraman et al. explored labeled data augmentation in one study\cite{rajaraman2020weakly}, achieving accuracies of 70.95\% and 88.89\% for two-class and three-class classifications and in another study\cite{9121222}, iteratively pruned ensembles with models like ResNet50 reaching up to 99.01\% accuracy. Lastly, Pir Masoom Shah\cite{9423965} combined CNN and GRU in a hybrid model, demonstrating high accuracy of 96\% on a smaller dataset of 424 images. These studies collectively highlighted the potential of DNNs in revolutionizing COVID-19 diagnosis through improved accuracy, data augmentation and innovative model architectures. 
\subsection{TL model developed to determine COVID-19} With the use of TL, COVID-19 recognition from CXR images has become significantly improved. The use of ResNet-50 and VGG-19 models in Roy's study \cite{roy2022svd} resulted in a 94\% accuracy rate. Using VGG-19, ResNet-50 and EfficientNet-B0, Monshi's CovidXrayNet optimized data augmentation and CNN hyperparameters achieving a 95.62\% accuracy rate. ResNet50 achieved a higher accuracy rate of 97.5\% on CT scans and 98.33\% on X-ray scans in comparison to VGG16 according to Berrimi's research \cite{9430229}. In his research incorporating ResNet50, InceptionV3, VGG16 and VGG19, Nefoussi\cite{9378703} emphasized that InceptionV3 achieved the greatest accuracy rate of 98.67\%. A robust COVID-19 identification approach was proposed by Asif's study \cite{9344870} on Inception V3 with TL which achieved over 98\% accuracy. ResNet-50 achieved 94\% accuracy on unseen data, thanks to fine-tuning advancements discovered in Ahmed's work into CNNs' generalization gap \cite{9430538}. With accuracies of 90.91\% and 94.44\%, respectively, Hernandez's study \cite{9311372} showed that ResNet-50 and VGG-16 performed better. The CNNs (VGG16, DenseNet201, ResNet50, and EfficientNetB3) were tested using LIME by Cervantes \cite{9569029}. The results showed that DenseNet201 and VGG16 achieved better accuracy in identifying COVID-19, healthy and viral pneumonia CXR pictures compared to ResNet50 and EfficientNetB3. These findings show that TL models work well to improve COVID-19 detection from CXR images, which is important for making better diagnostic tools.

The presented research examined the identification of COVID-19 in CXR images utilizing ML, DL and TL techniques. While the results are promising, limitations include small and unbalanced datasets, restricted applicability to larger populations, data inaccuracies, lack of clinical validation and difficulty in fully explaining AI decisions. Utilizing Bangladeshi CXR data and applying SMOTE to correct class imbalances, this study evaluates ML models and incorporates XAI to enhance decision-making transparency aiming to create a more accurate, reliable and interpretable detection system.

\section{Dataset}
The Dataset is employed throughout this research is openly available at Mendeley Data\footnote{\url{https://data.mendeley.com/datasets/m4s2jn3csb/1}}. The collection consists of 4,350 high-quality CXR images collected from numerous medical institutions from Bangladesh. The images are categorized into four distinct groups: Normal (consisting of 1,200 photos), Lung-opacity (comprising 1,100 images), COVID-19 (consisting of 1,050 images) and Viral-pneumonia (comprising 1,000 images). These images are an invaluable resource for identifying and studying lung diseases like COVID-19. The collection enables prompt identification and monitoring of these diseases, which represent substantial worldwide health issues. By employing ML and DL techniques, this dataset can improve patient care, decrease mortality rates linked to COVID-19 and respiratory diseases, and aid in the creation of automated diagnostic instruments.
\section{Methodology}
\subsection{Data Preprocessing}The data preprocessing has enhanced CXR images for subsequent analysis by applying several image processing techniques. Each image is resized to 256x256 pixels. The dataset is divided with 80\% for training, 10\% for testing, and 10\% for validation. Then the following steps are applied:

\begin{enumerate}[label=\roman*]
    \item Resize and convert to Grayscale.
    \item Apply Laplacian filter: \(\nabla^2 I\).
    \item Sharpen image with kernel \(K\): \(S = \text{sharpen}(I)\); subtract Laplacian-filtered image from sharpened image: \(B = S - L\).
    \item Apply Sobel filters: \(G_x = \frac{\partial I}{\partial x}\), \(G_y = \frac{\partial I}{\partial y}\), and combine them: \(G = \sqrt{G_x^2 + G_y^2}\).
    \item Create mask: \(M = G \cdot B\).
    \item Combine mask with Laplacian-filtered image: \(F = L + M\).
    \item Apply power-law transformation: \(I' = 255 \cdot \left( \frac{I}{255} \right)^\gamma\).
\end{enumerate}

\begin{figure}[ht]
    \centering
    \begin{subfigure}[b]{0.9\linewidth}
        \centering
        \begin{subfigure}[b]{0.2\linewidth}
            \includegraphics[width=\linewidth]{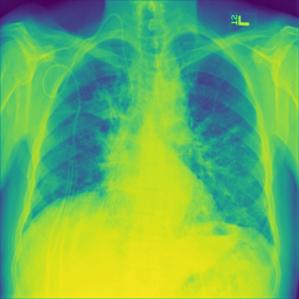}
        \end{subfigure}
        \hfill
        \begin{subfigure}[b]{0.2\linewidth}
            \includegraphics[width=\linewidth]{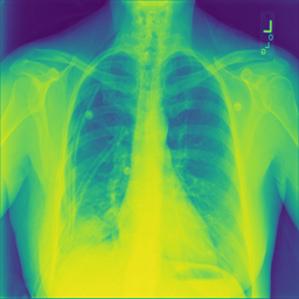}
        \end{subfigure}
        \hfill
        \begin{subfigure}[b]{0.2\linewidth}
            \includegraphics[width=\linewidth]{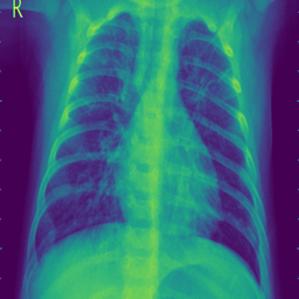}
        \end{subfigure}
        \hfill
        \begin{subfigure}[b]{0.2\linewidth}
            \includegraphics[width=\linewidth]{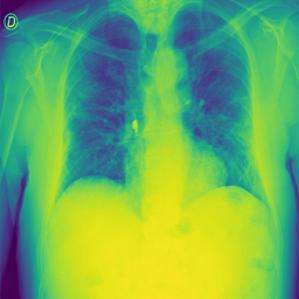}
        \end{subfigure}
        \caption{Before Preprocessing}
    \end{subfigure}
    
    \vskip\baselineskip
    
    \begin{subfigure}[b]{0.9\linewidth}
        \centering
        \begin{subfigure}[b]{0.2\linewidth}
            \includegraphics[width=\linewidth]{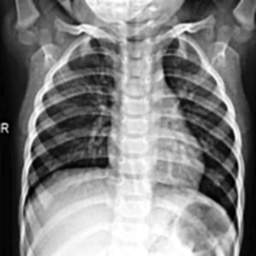}
        \end{subfigure}
        \hfill
        \begin{subfigure}[b]{0.2\linewidth}
            \includegraphics[width=\linewidth]{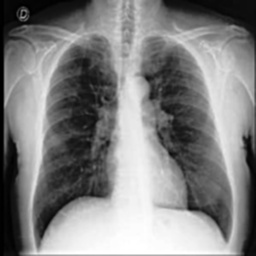}
        \end{subfigure}
        \hfill
        \begin{subfigure}[b]{0.2\linewidth}
            \includegraphics[width=\linewidth]{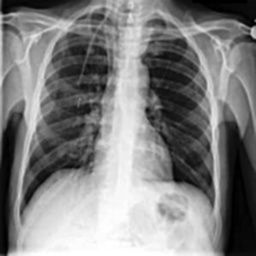}
        \end{subfigure}
    \hfill
        \begin{subfigure}[b]{0.2\linewidth}
            \includegraphics[width=\linewidth]{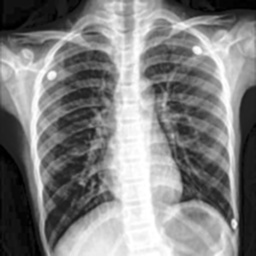}
        \end{subfigure}
    \caption{After Preprocessing}
    \end{subfigure}
    \caption{Comparison of CXR Images Before and After Preprocessing}
    \label{Figure 1}
\end{figure}

These steps collectively enhance the critical features of the CXR images, aiding in more accurate analysis. Figure\ref{Figure 1} shows the comparison between CXR images before and after preprocessing highlighting the enhanced clarity and feature extraction achieved through the applied filters.

\subsection{SMOTE}
SMOTE\cite{chawla2002smote} which is an over-sampling technique utilized to address class imbalance by generating synthetic instances rather than duplicating existing ones. Operating in the "feature space," SMOTE creates artificial instances to balance the dataset. The \texttt{sampling\_strategy} parameter defines the desired ratio of minority to majority class samples and is configured as a dictionary mapping each class label to the target number of samples. Other parameters like \texttt{random\_state}, \texttt{k\_neighbors} and \texttt{n\_jobs} are left at their default values but can be adjusted for optimization. Our SMOTE implementation follows a two-phase approach: first, I oversampled the minority class to achieve balance; second, I applied the generated synthetic instances to the dataset, improving class representation and mitigating imbalance issues\cite{JMLR:v18:16-365}. My implementation of SMOTE utilizes a two-phase methodology:
\begin{enumerate}[label=\roman*.]
  \item \textbf{SMOTE-1}: Firstly, I have applied SMOTE to balance the classes with equal 1200 samples.
  \item \textbf{SMOTE-2}: Then, I have applied SMOTE to balance the classes with equal 1500 samples.
\end{enumerate}
I have compared the analysis findings using varying sample sizes per class to evaluate how oversampling impacts the model's performance and accuracy.
\begin{figure*}[ht]
    \centering
   \includegraphics[width=\linewidth]{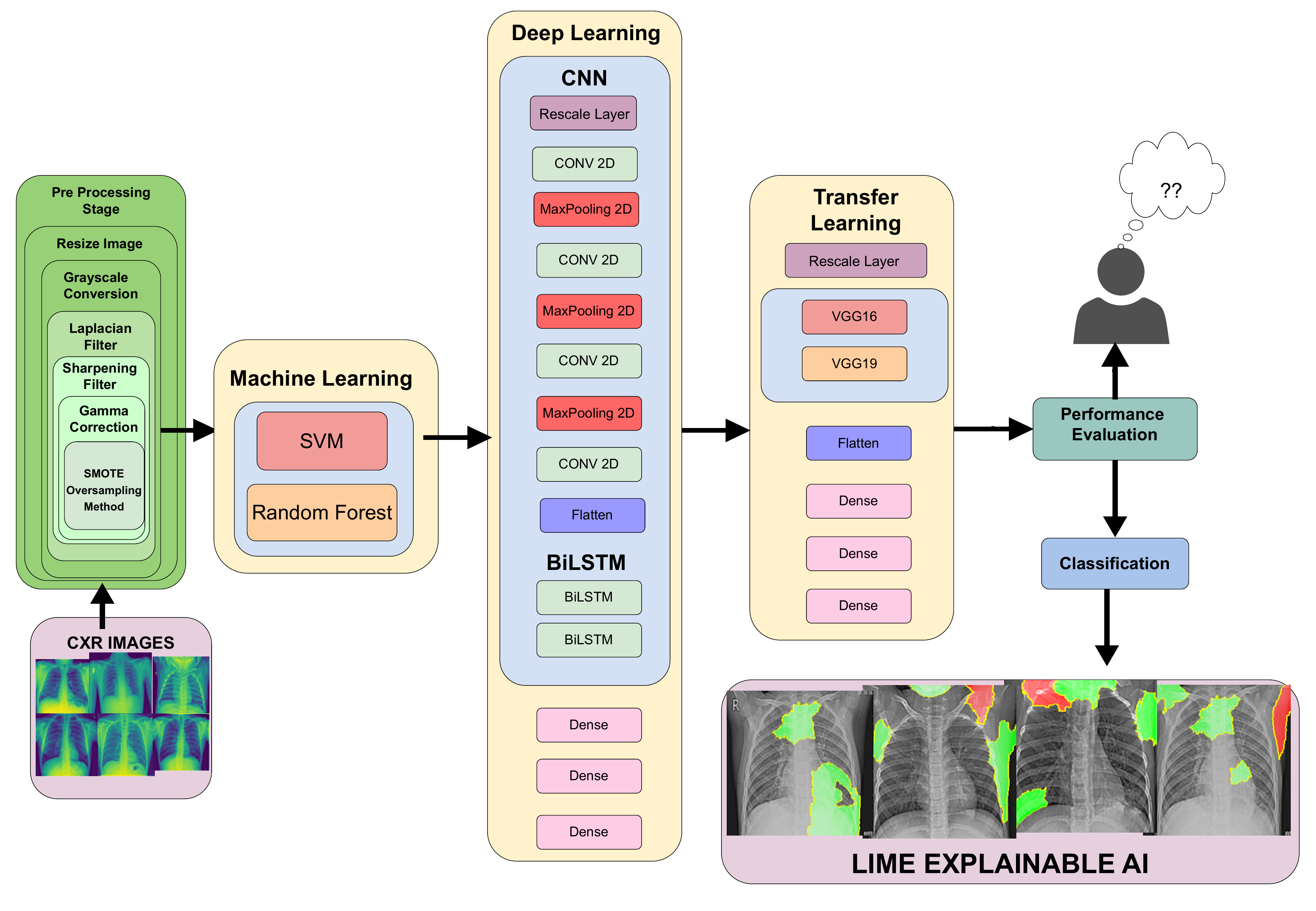}
    \caption{Proposed Model Architecture}
    \label{fig:proposed_model}
\end{figure*}
\subsection{Model Architecture}
The proposed framework for this research is a combination structure that integrates both traditional ML and DL techniques to diagnose COVID-19 from CXR data. Figure\ref{fig:proposed_model} illustrates the suggested model's structure. Firstly, CXR images undergo pre-processing, which includes converting them to grayscale, performing Laplacian filtering, sharpening, and gamma correction. Next, the images are reduced in size and the dataset is divided. The pre-processed images are rescaled and then inputted into a CNN consisting of multiple layers: three convolutional layers, each followed by max-pooling, a flattening layer, and dense layers for feature extraction.

Furthermore, the architecture incorporates pre-trained models like VGG16 and VGG19, to which additional dense layers were added for better performance. These models are combined with a BiLSTM layer for sequential learning. HOG is used to extract features from the CXR images, and these features are subsequently utilized to train machine learning models, including SVM and Random Forest (RF). The proposed framework leverages CNN and BiLSTM as DL methods and integrates SVM and RF as ML models to classify the CXR images. SMOTE is utilized to tackle the problem of inequalities between classes. The model includes a classification layer and employs LIME to offer interpretable AI, hence ensuring the interpretability of the model's judgments. Performance evaluation is carried out by utilizing metrics to gauge the precision and dependability of the model.

\section{Result Analysis}
 Table\ref{tab:performance_comparison} presents a detailed comparison of model performance across three dataset condition: the Original Dataset, Smote 1, and Smote 2. In the Original one, the highest result is achieved by VGG 16 and VGG 19 models, both scoring \textbf{92\%}, demonstrating their superior performance among the models tested. In the Smote 1 dataset, the VGG 16 model has acheived with an accuracy of \textbf{95\%}, while VGG 19 followed closely with \textbf{94\%}. The Smote 2 dataset has showed the highest accuracy overall with VGG 19 achieving an impressive \textbf{98\%}, and VGG 16 also performing strongly at \textbf{96\%}. ML models like SVM and Random Forest showed lower accuracies across all datasets compared toTL models. Similarly, DL models such as CNN and BiLSTM have performed well but are still outperformed by VGG models. This trend highlights the effectiveness of TL, particularly VGG architectures, in handling the datasets. Additionally, while oversampling with techniques like SMOTE generally enhances model performance by addressing class imbalances, excessive oversampling can cause overfitting. This is evident as the Smote 2 dataset, with a balanced approach to oversampling, produced the best results. Therefore, finding the right level of oversampling is crucial to optimizing model performance without inducing overfitting.

\begin{table}[!ht]
\renewcommand{\arraystretch}{1.7}
\centering
\caption{Model Performance Comparison}
\begin{tabular}{|c|c|c|c|c|c|c|}
\hline
\textbf{Dataset}                   & \textbf{\shortstack{Model\\type}} & \textbf{Model}         & \textbf{Accuracy} & \textbf{Precision} & \textbf{Recall} & \textbf{F1 Score} \\ \hline
\multirow{6}{*}{\shortstack{Original Dataset}} & \multirow{2}{*}{\shortstack{Machine Learning}}  & SVM           & 0.84        & 0.84         & 0.84      & 0.84        \\ \cline{3-7} 
                          &                                    & \shortstack{Random Forest} & 0.78        & 0.78         & 0.78      & 0.78        \\ \cline{2-7} 
                          & \multirow{2}{*}{\shortstack{Deep Learning}}     & CNN           & 0.91        & 0.93         & 0.92      & 0.92        \\ \cline{3-7} 
                          &                                    & BiLSTM        & 0.75        & 0.77         & 0.76      & 0.77        \\ \cline{2-7} 
                          & \multirow{2}{*}{\shortstack{Transfer Learning}} & \textbf{VGG-16}        & \textbf{0.92}        & \textbf{0.92}         & \textbf{0.92}      & \textbf{0.92}        \\ \cline{3-7} 
                          &                                    & \textbf{VGG-19}        & \textbf{0.92}        & \textbf{0.92}         & \textbf{0.92}      & \textbf{0.92}        \\ \hline
\multirow{6}{*}{Smote 1}  & \multirow{2}{*}{\shortstack{Machine Learning}}  & SVM           & 0.85        & 0.85         & 0.85      & 0.85        \\ \cline{3-7} 
                          &                                    & \shortstack{Random Forest}  & 0.78        & 0.77         &0.77      & 0.77        \\ \cline{2-7} 
                          & \multirow{2}{*}{\shortstack{Deep Learning}}     & CNN           & 0.92        & 0.93         & 0.92      & 0.92        \\ \cline{3-7} 
                          &                                    & BiLSTM        & 0.77        & 0.77         & 0.76      & 0.76        \\ \cline{2-7} 
                          & \multirow{2}{*}{\shortstack{Transfer Learning}} & \textbf{VGG-16 }       & \textbf{0.95 }      & \textbf{0.96}    & \textbf{0.96}       & \textbf{0.96}        \\ \cline{3-7} 
                          &                                    & VGG-19        & 0.94        & 0.94         & 0.94      & 0.94        \\ \hline
\multirow{6}{*}{Smote 2}  & \multirow{2}{*}{\shortstack{Machine Learning}}  & SVM           & 0.90        & 0.90         & 0.89      &0.89         \\ \cline{3-7} 
                          &                                    & \shortstack{Random Forest}  & 0.85        & 0.84         & 0.84      & 0.84        \\ \cline{2-7} 
                          & \multirow{2}{*}{\shortstack{Deep Learning}}     & CNN           & 0.94        & 0.95         & 0.94      & 0.94        \\ \cline{3-7} 
                          &                                    & BiLSTM        & 0.84        & 0.84         & 0.83      & 0.83        \\ \cline{2-7} 
                          & \multirow{2}{*}{\shortstack{Transfer Learning}} & VGG-16        & 0.96        & 0.97         & 0.95      & 0.96        \\ \cline{3-7} 
                          &                                    & \textbf{VGG-19}        & \textbf{0.98}        & \textbf{0.99}         &\textbf{ 0.99}      & \textbf{0.98}        \\ \hline
\end{tabular}
\label{tab:performance_comparison}
\end{table}

\section{Explainable AI}
\begin{figure}[ht!]
    \centering
    \begin{subfigure}[b]{0.8\linewidth}
        \includegraphics[width=\linewidth]{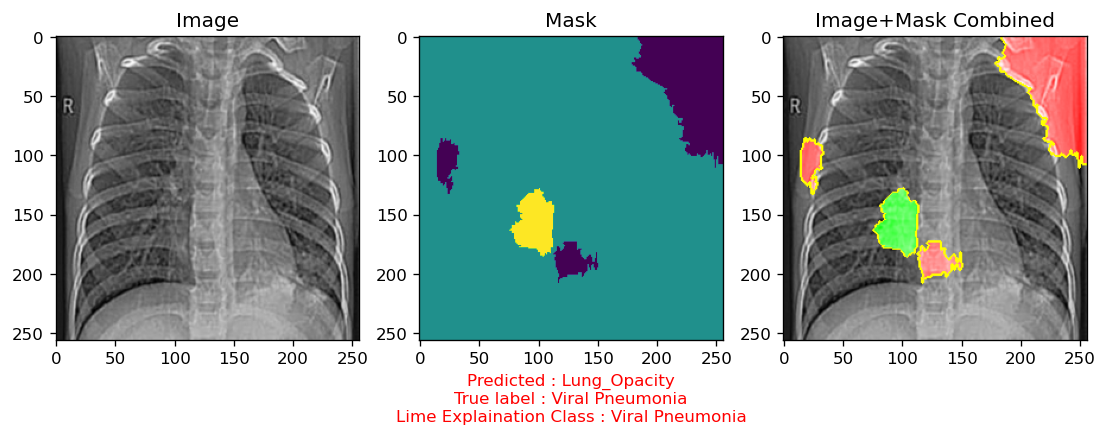}
        \captionsetup{justification=centering}
    \end{subfigure}
    \begin{subfigure}[b]{0.8\linewidth}
        \includegraphics[width=\linewidth]{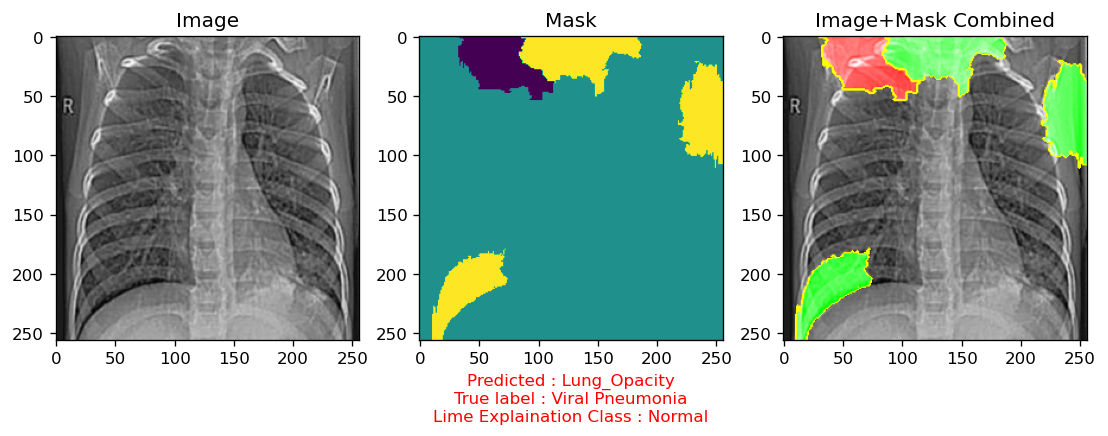}
        \captionsetup{justification=centering}
    \end{subfigure}
    \begin{subfigure}[b]{0.8\linewidth}
        \includegraphics[width=\linewidth]{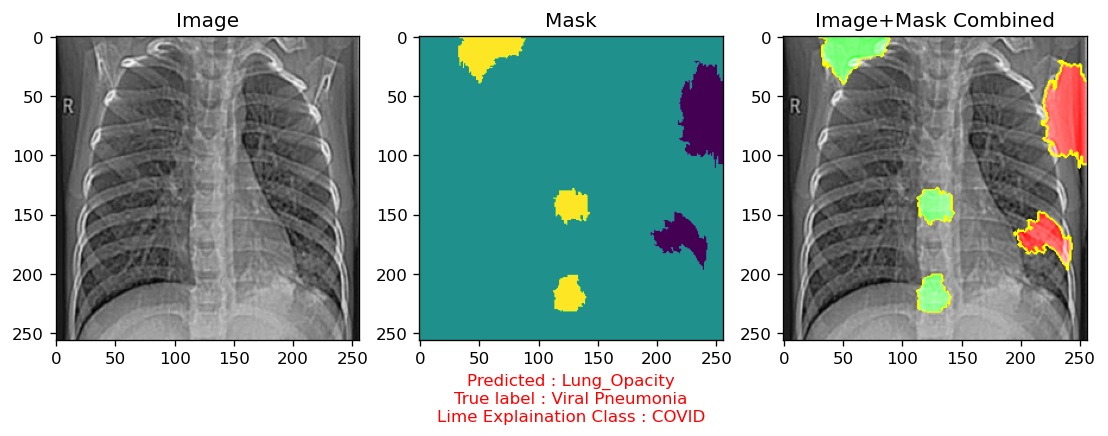}
        \captionsetup{justification=centering}
    \end{subfigure}
    \begin{subfigure}[b]{0.8\linewidth}
        \includegraphics[width=\linewidth]{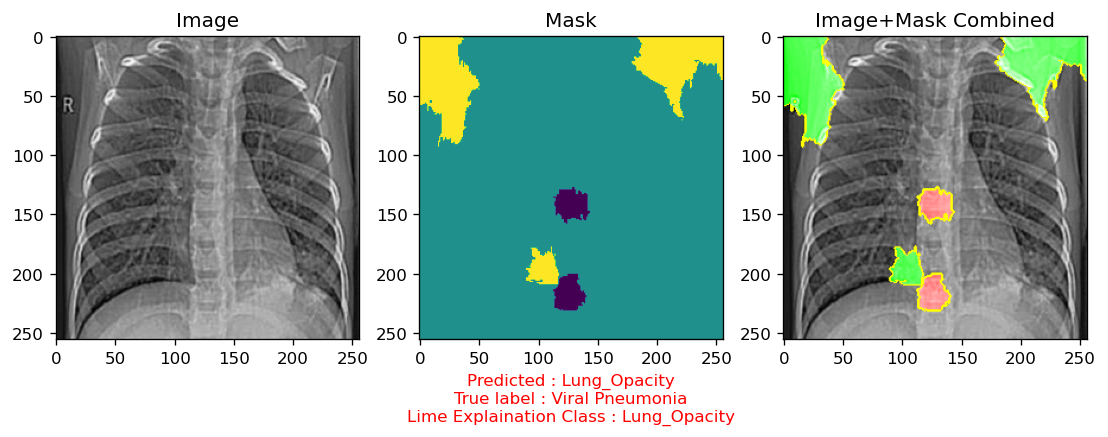}
        \captionsetup{justification=centering}
    \end{subfigure}
    \caption{ Some Sample Images of Actual Class Predicted Incorrectly by TL Model}
    \label{Figure 3}
\end{figure}

Neural network models are often described as inscrutable systems because understanding the reasoning behind their predictions is challenging. This is where Explainable AI (XAI) becomes important; XAI helps us understand how these models make decisions. In this study, LIME as our explainable AI method. created by Ribeiro et al\cite{ribeiro2016should}. LIME can explain how any classifier or regressor works, no matter how complex. It does this by approximating the model locally to something simpler and easier to understand, making the reasons behind the model's predictions clear. This way, LIME helps bridge the gap between complicated machine learning models and the people using them, providing valuable insights into why the models make certain decisions.
Figure\ref{Figure 3} shows how LIME explains a model's misclassifications of chest X-rays. \textit{Lung Opacity} is the actual label; \textit{Viral Pneumonia} is what the model predicted. LIME indicated in green the regions with \textit{Viral Pneumonia} and in red the other classes in the first plot. The second set of plot displays LIME highlighting in green the regions that belong to \textit{Normal} and in red other classes. LIME highlights in the third set regions that indicate \textit{Covid} in green and other classes in red. LIME indicated in the fourth set the regions that correspond to \textit{Lung Opacity} in green and the other classes in red.

LIME's predictions can be understood by examining the highlighted areas in green and red. The green areas represent the actual regions for which the model predicts the given class, while the red areas correspond to other classes. In the fourth plot, the green area for \textit{Lung Opacity} is the largest compared to the green areas in the other three plots. This is why the model predicted \textit{Lung Opacity}, even though the true label is \textit{Viral Pneumonia}. The larger green area indicates that the model's decision is primarily influenced by these regions, showing that the model focuses more on features associated with \textit{Lung Opacity}. By analyzing these highlighted areas, we can understand why the model made its predictions, revealing the reasons behind the black box of the model's predictability and helping identify potential misclassifications. This analysis helps us see how the model might be misled by certain features, emphasizing the importance of examining these explanations to improve model accuracy and reliability.

\begin{figure}[ht!]
    \centering
    \begin{subfigure}[b]{0.8\linewidth}
        \includegraphics[width=\linewidth]{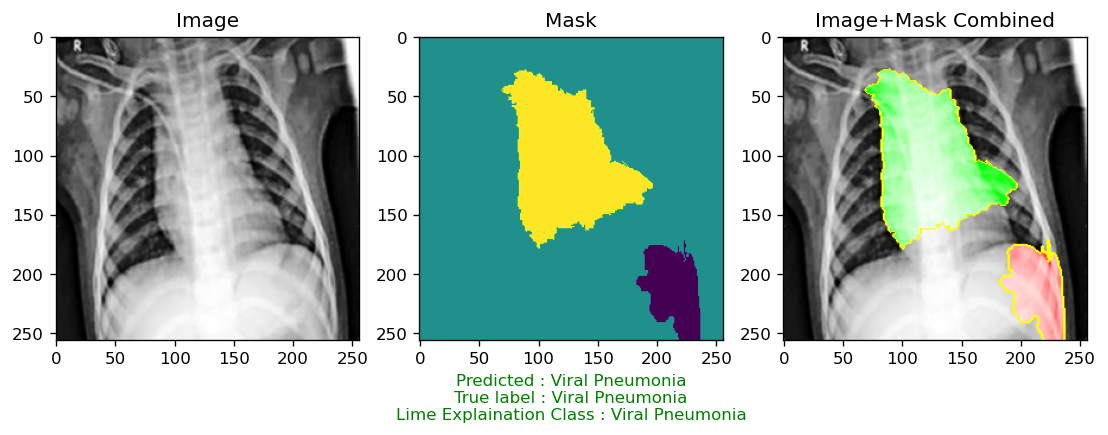}
        \captionsetup{justification=centering}
    \end{subfigure}
    \begin{subfigure}[b]{0.8\linewidth}
        \includegraphics[width=\linewidth]{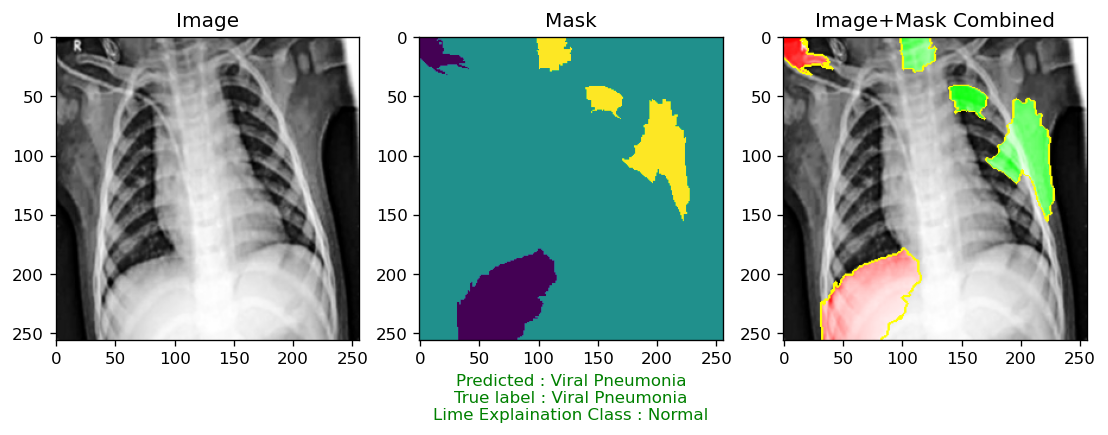}
        \captionsetup{justification=centering}
    \end{subfigure}
    \begin{subfigure}[b]{0.8\linewidth}
        \includegraphics[width=\linewidth]{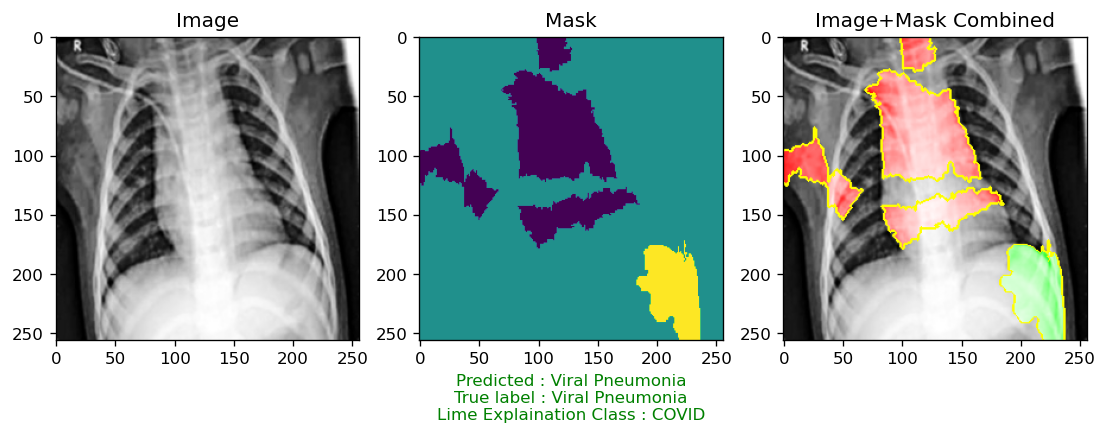}
        \captionsetup{justification=centering}
    \end{subfigure}
    \begin{subfigure}[b]{0.8\linewidth}
        \includegraphics[width=\linewidth]{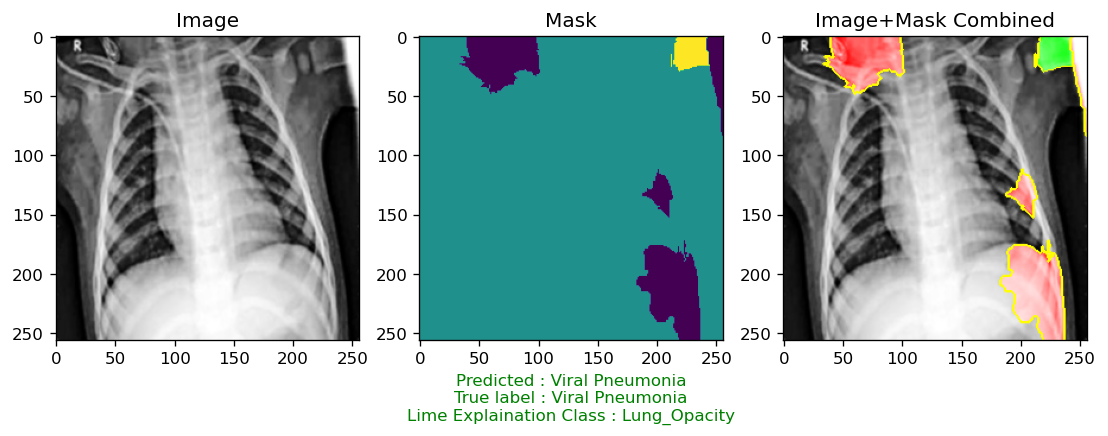}
        \captionsetup{justification=centering}
    \end{subfigure}
    \caption{ Some Sample Images of Actual Class Predicted correctly by TL Model}
    \label{Figure 4}
\end{figure}
The Figure\ref{Figure 4} demonstrates how LIME explains the model's correctly classified prediction of a chest X-ray. The true label is \textit{Viral Pneumonia} and the model predicted \textit{Viral Pneumonia}, correctly classifying the image. In the first set, LIME highlighted areas corresponding to \textit{Viral Pneumonia} in green, while marking other classes in red. In the second set, LIME highlighted areas suggesting \textit{Normal} in green, with other classes in red. In the third set, LIME highlighted regions indicating \textit{COVID} in green, and other classes in red. In the fourth set, LIME highlighted areas associated with \textit{Lung Opacity} in green, with other classes in red.LIME functions by graphically emphasizing the particular areas inside an image that have substantially influenced the model's prediction.

In the initial image, the green area for \textit{Viral Pneumonia} is larger than the green areas in the other three images. This is why the model is influenced by these features and predicted \textit{Viral Pneumonia}. This approach helps maintain transparency between the model and its predictions, allowing us to understand the specific reasons behind the model's decisions. By examining these highlighted areas, we can see how the model is influenced by certain features, ensuring that we know exactly why the model predicted a particular class. This transparency is crucial for validating the model's accuracy and reliability.

XAI is crucial for both misclassifications and correct classifications. It provides transparency into machine learning models' decisions, enabling us to identify and correct potential errors. XAI reveals why a model is misled, guiding improvements. It ensures valid features in correct classifications, boosting confidence in predictions and enhancing trust in the AI system.
\section{Conclusion \& Future Work}
In conlusion, this study successfully applies various ML and DL algorithms to accurately detect abnormalities from CXR images, with a particular focus on data from Bangladesh. By using SMOTE to address class imbalances and implementing XAI techniques like LIME to enhance model interpretability, the study improves both the precision and reliability of detection.The VGG19 model attains the utmost performance, showcasing the efficacy of DL. Future research should focus on a more detailed analysis of models to optimize their performance. Additionally, expanding the dataset with more images from Bangladesh is essential for improving model reliability and robustness. Further development of XAI techniques is also necessary to ensure transparency and trust in automated healthcare systems.

\bibliographystyle{unsrt} 
\bibliography{references.bib}

\end{document}